\documentclass{svjour3}                     % onecolumn (standard format)
\smartqed  % flush right qed marks, e.g. at end of proof
\usepackage{graphicx}
\usepackage{wrapfig}

\journalname{Few-Body Systems (FB20)}
\begin{document}

\title{
%Insert your title here
Hadronic few-body systems in chiral dynamics 
%\thanks{This work was partially supported by the Grants-in-Aid for Scientific Research (No. 24105706). This work was done in part under the Yukawa International Program for Quark- hadron Sciences (YIPQS).}
}
\subtitle{
Few-body systems in hadron physics
%Subtitle
}


\author{Daisuke Jido  
%       \and
%        Second Author %etc.
}


\institute{D. Jido \at
              Yukawa Institute for Theoretical Physics
              Kyoto University, Kyoto 606-8502, JAPAN\\
              Tel.: +81-75-7537034\\
              Fax: +81-75-7537010\\
              \email{jido@yukawa.kyoto-u.ac.jp}           %  \\
}

\date{Received: date / Accepted: date}

\maketitle

\begin{abstract}
Hadronic composite states are introduced as few-body systems in hadron physics. The $\Lambda(1405)$ resonance is a good example of the hadronic few-body systems. It has turned out that $\Lambda(1405)$ can be described by hadronic dynamics in a modern technology which incorporates coupled channel unitarity framework and chiral dynamics. The idea of the hadronic $\bar KN$ composite state of $\Lambda(1405)$ is extended to kaonic few-body states. It is concluded that, due to the fact that $K$ and $N$ have similar interaction nature in $s$-wave $\bar K$ couplings, there are few-body quasibound states with kaons systematically just below the break-up thresholds, like $\bar KNN$, $\bar KKN$ and $\bar KKK$, as well as $\Lambda(1405)$ as a $\bar KN$ quasibound state and $f_{0}(980)$ and $a_{0}(980)$ as $\bar KK$. 
\keywords{Hadronic composite state \and $\Lambda(1405)$ resonance \and 
kaonic few-body systems \and chiral dynamics}
\end{abstract}

\section{Introduction}
\label{intro}

Hadrons are composite objects of quark and gluons governed by quantum chromodynamics, QCD. So far hundreds of hadrons have been experimentally observed in spite of only the five (or six) fundamental pieces, up, down, strange, charm, bottom (and top) quarks. The richness of hadron spectrum is a consequence of highly nontrivial dynamics of quarks and gluons confined in hadron. Precise measurement of hadron spectrum is one of the phenomenological ways to view colored dynamics of quarks and gluons inside hadron, and gives us clues of the  confinement mechanism. Especially the systematics of energy excitation is a key issue to understand the hadron structure. 

A conventional picture of hadron is the quark model, in which one considers constituent quarks as quasi-particles moving in a one-body mean field created by gluon dynamics, and symmetry of the quasi-particles and (weak) residual interaction characterize the hadron structure. Although quark models work well for lowest-lying states, there are a lot of states found which do not match quark model predictions, which are so-called exotic hadrons, and quark models should incorporate hadron dynamics once strong decay channels open. It has been realized that quark models do not give us the universal picture of the hadron structure, and that the constituent quark alone cannot be an effective degree of freedom in hadron.  These facts brings us the idea to consider other effective constituents, such as diquark correlation, multiquark component and hadronic composite, which will be the origin of  richness of  hadron spectrum. 

\section{Hadronic composite states}
\label{sec:1}

Hadron composite state can be one of the existence forms of hadron. Its constituents are hadrons governed by hadronic dynamics not inter-quark colored force. Consequently the spacial size is much larger than the typical hadron size which is characterized by the confining force. The typical examples are atomic nuclei which are bound states of nucleons. Mesons also can be constituents of the hadron composite states. Because of no number conservation of mesons, the composite state with mesons has transition modes to lighter mesons, mostly pions, and absorptive decay modes. Therefore, most of hadronic composite states are unstable bound states with decay modes. 

Hadron composite state can be a new sample to investigate the confinement mechanism. Most of hadron flavors can be explained by a quark-anitquark pair $\bar qq$ and three quarks $qqq$. So far, very few flavor exotic hadrons have been observed. This is an important question to be solved in hadron physics. 
A clue is multiquark state. 
Multiquark state always has color singlet clusters. Thus, there is competition between colorful hadron {\it constituent force} inside confinement range and colorless hadron {\it interaction force} in a larger ragne than the typical hadron size. For instance, the $H$-dibaryon composed by $uuddss$ can have a colored three diquark configuration $([ud][ds][su])$ governed by short range color interaction and a colorless two baryon configuration $([\Lambda\Lambda]$ and/or $[\Xi N])$ governed by longer range hadronic interaction. The real state can be a mixture of these configuration. Nevertheless, since the former configuration should be a compact object, while the latter can be a larger object, there is a (small) scale gap between two configurations. These two have also different systematics in excitation modes and flavor symmetry. Thus, it would be very interesting if one could know which systematics is realized in each hadron. 

\section{The $\Lambda(1405)$ resonance}
\label{sec:2}

$\Lambda(1405)$ is a baryon resonance with isospin $I=0$ and strangeness $S=-1$ sitting between the $\bar KN$ (1435 MeV) and $\pi\Sigma$ (1331 MeV) thresholds with  mass around 1405 MeV and 50 MeV decay width to $\pi\Sigma$. The flavor of $\Lambda(1405)$ can be expressed by the minimal quark contents of $uds$, however, simple quark models have failed to reproduce the $\Lambda(1405)$ mass. 
$\Lambda(1405)$ is a historical example of the hadronic composite state.  $\Lambda(1405)$  has been considered as a quasibound state of $\bar KN$~\cite{Dalitz:1959dn,Dalitz:1960du}, even before QCD was established. In general, the physically observed state should be given as a mixture of the hadronic composite and quark model type components, it is important to examine the dominant components of hadron resonance states to understand their structure and dynamics, and it would be interesting if one could know the fraction of the components. 

For the theoretical description of $\Lambda(1405)$, one needs dynamical study  of coupled channels including at least $\bar KN$ and $\pi\Sigma$, because the $\Lambda(1405)$ resonance is located in the 100 MeV window of the $\bar KN$ and $\pi\Sigma$ thresholds.  Flavor SU(3) complete treatment of coupled channels, by including $\pi \Sigma$, $\bar KN$, $\pi\Lambda$, $\eta \Lambda$, $\eta \Sigma$ and $K \Xi$, was already done using phenomenological vector-meson exchange potential in Ref.~\cite{Dalitz:1967fp}. Coupled channels calculation of $\pi\Sigma$ and $\bar KN$ based on the cloudy bag model was done in Ref.~\cite{Veit:1984jr}\footnote{In this model, a bare pole term for the $\Lambda(1405)$ resonance  was explicitly introduced around 1650 MeV.}. The modern approach based on chiral dynamics and unitary coupled channels was initiated by Ref.~\cite{Kaiser:1995eg}, and the $\Lambda(1405)$ resonance is well described as well as the $K^{-}p$ threshold properties and $K^{-}p$ scatterings. (See Ref.~\cite{Hyodo:2011ur} as a recent review article and Ref.~\cite{Ikeda:2011pi,Ikeda:2012au} for a very recent update including the new SIDDHARTA measurement~\cite{Bazzi:2011zj}.)

One of the most important consequences of the coupled channels approach based on chiral dynamics is that the $\Lambda(1405)$ resonance is a superposition of two states~
\cite{Oller:2000fj,Jido:2003cb}. These two states have different properties~\cite{Jido:2003cb}: one of the states is located at 1426 MeV with a 32 MeV width and couples dominantly to $\bar KN$, while the other is at 1390 MeV with a broader width and couples strongly to $\pi\Sigma$. All of the recent calculations based on chiral interaction suggest the two pole structure of  $\Lambda(1405)$  although the pole positions are model dependent slightly. The reason that there are two states around the $\Lambda(1405)$ energy region is that the chiral interaction indicates two attractions with $I=0$ in the flavor SU(3) singlet and octet channels group-theoretically~\cite{Jido:2003cb} or in the $\bar KN$ and $\pi\Sigma$ channels physically~\cite{Hyodo:2007jq}. The latter fact implies that the $\Lambda(1405)$ is essentially described by two dynamical channels of $\bar KN$ and $\pi\Sigma$, and that  the higher state is a $\bar KN$ quasibound state decaying to $\pi\Sigma$ and the lower is a $\pi\Sigma$ resonant state~\cite{Hyodo:2007jq}. As a consequence of the double pole nature, the $\Lambda(1405)$ spectrum depends on the initial channel~\cite{Jido:2003cb}. Thus, observing that the $\Lambda(1405)$ spectrum in the $\bar KN \to \pi\Sigma$ channel is different from that in  $\pi\Sigma \to \pi\Sigma$, one can confirm that $\Lambda(1405)$ is a dynamical object of $\bar KN$ and $\pi\Sigma$. For this purpose, since $\Lambda(1405)$ is located below the threshold of $\bar KN$, one needs indirect production of $\Lambda(1405)$ from $\bar KN$. It has been found in Refs.~\cite{Jido:2009jf,Jido:2010rx} that in the $K^{-}d \to \Lambda(1405) n$ reaction $\Lambda(1405)$ is selectively produced by $\bar KN$, thus this reaction is one of the suitable reactions to investigate the nature of $\Lambda(1405)$. An experiment of this reaction at J-PARC is proposed~\cite{Noumi} observing neutrons in the forward direction.

\section{Compositeness of hadron}
Compositeness of hadron had been discussed before QCD was established in order to find fundamental or elementary hadrons. But this trial was certainly failed, since all of the hadrons are composite objects of quarks and gluons. Here we discuss the compositeness of hadron resonances in terms of quark originated state and hadronic composite state. To discuss the compositeness of the hadron resonances, first of all, we need to define elementary hadrons. The elementary hadrons may be the ground state hadrons stable against strong interactions, or one may take hadrons which survive in the large $N_{c}$ limit. Once one defines the elementary hadrons, one can discuss the compositeness of hadron quantitatively. The elementary components are expressed in the free Hamiltonian $H_{0}$ and the composite states are dynamically generated as a consequence of the interaction of the elementary components $V$ out of the full Hamiltonian of the system $H=H_{0}+V$. The compositeness index can be written as $1-Z$ where $Z = \sum_{n} | \langle n | d \rangle |^{2}$ with the eigenstates $|n\rangle $ of $H_{0}$ and the dynamically generated state $|d \rangle$ as an eigenstate of $H$~\cite{Weinberg:1965aa}. This definition of the compositeness can be extended in field theory by introduction the Lagrangian of the system as $L=L_{0}+L_{\rm int}$ in which $L_{0}$ is the free Lagrangian of the elementary component and $L_{\rm int}$ represents their interaction. The field renormalization constant $Z$ is shown in a reside of the full propagator $\Delta$ satisfying Dyson equation at the pole position, of which normalization is given by the free propagator $\Delta_{0}(E) = 1/(E-M_{0})$ with the bare mass $M_{0}$~\cite{PhysRev.136.B816}. An application for chiral dynamics is discussed in Ref.~\cite{Hyodo:2011qc}. For the resonance state, the compositeness index can be a complex number. Thus, one needs its appropriate interpretation, which is an open issue yet.

Here we would like to discuss the compositeness  of  $\Lambda(1405)$ in a different way within the chiral unitary approach~\cite{Hyodo:2008xr}. In the chiral unitary approach, one solves Lippmann-Schwinger equation $T=V+VGT$, in which $V$ is the interaction kernel, while $G$ is the loop function which guarantees unitarity and specifies the model space. Thus, in the chiral unitary approach, the elementary components are given in the loop function. In the present case for $\Lambda(1405)$, they are the lowest lying octet baryons and mesons. The interactions among the elementary components are given by the chiral effective theory. In the context of the discussion of compositeness, one has to take care of the interaction kernel, since in the interaction kernel sources of the resonances can be hidden. It is well-known that the $s$-channel resonance contributions are involved in the contact interactions of the elementary component~\cite{Ecker:1988te}. Nevertheless, the Weinberg-Tomozawa interaction may be free from the $s$-wave resonance contributions since it comes from the $t$-channel vector meson exchange. For the discussion of the compositeness of the resonance, let us take only the Weinberg-Tomozawa interaction as the interaction kernel. Since the Weinberg-Tomozawa interaction is a short range contact force, one needs to regularize the loop integral. In the regularization procedure, one fixes high-momentum behavior which is not controlled in the present model space. Therefore, even though one takes only hadronic interaction in interaction kernel, some contributions coming from outside of the model space can be hidden in the regularization parameters~\cite{Hyodo:2008xr}. In the chiral unitary model, this parameter is fixed phenomenologically so as to reproduce observed scattering cross section. Thus, if necessary, nature will request the components which come from the outside of the model space, namely, quark originated component.

\begin{figure}
\centering 
  \includegraphics[width=0.9\textwidth,bb=0 0 1003 345]{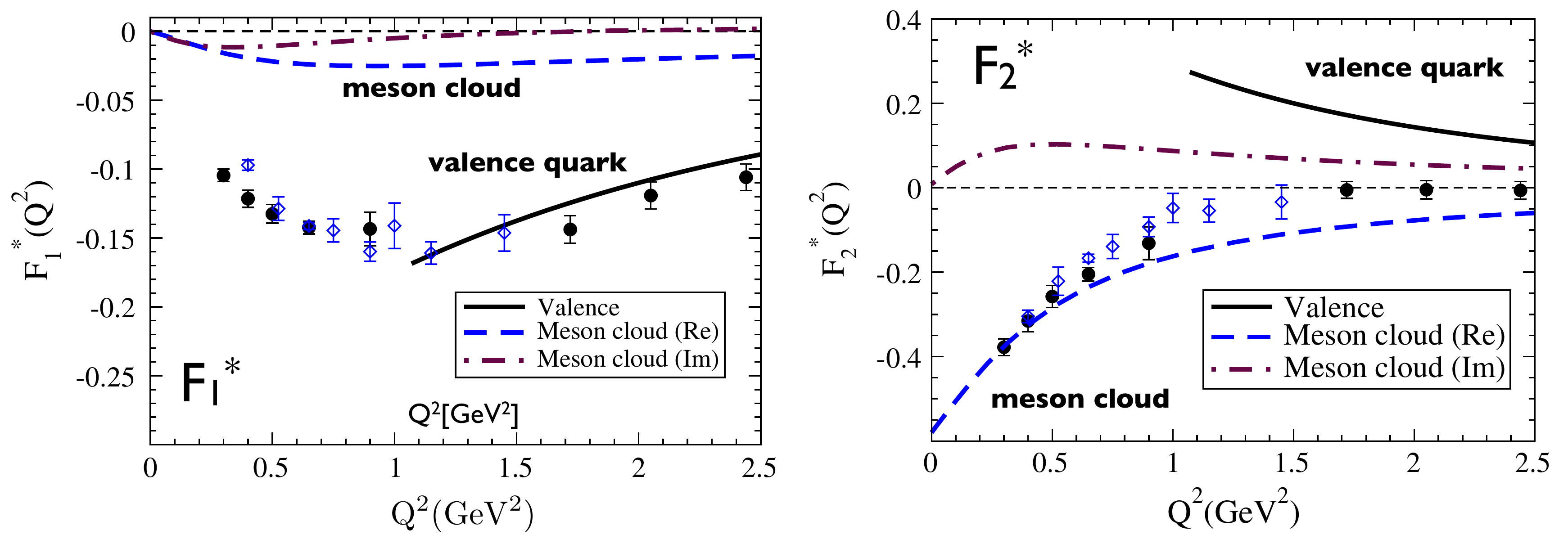}
  \caption{Transition form factors of $\gamma^{*} p \to N(1535)$ calculated by the chiral unitary model (meson cloud)~\cite{Jido:2007sm} and spectator quark model (valence quark)~\cite{Ramalho:2011ae,Ramalho:2011fa}. The plots are taken from Ref. ~\cite{Ramalho:2012im}.}
  \label{fig:FF}
\end{figure}

We have a choice of the parameter in which the hidden contribution can be excluded from formulation by theoretical requirement on the renormalization constant (natural renormalization scheme)~\cite{Hyodo:2008xr}.
If resonances can be reproduced by the chiral unitary approach with 
the Weinberg-Tomozawa interaction in the natural renormalization scheme, 
the resonances can be regarded as composite objects of meson and baryon 
constituents. In Ref.~\cite{Hyodo:2008xr} it has been found that the pole positions
of $\Lambda(1405)$ is well reproduced by the natural renormalization scheme,
while $N(1535)$ is not. This suggests that $\Lambda(1405)$ is mostly a composite
state of the ground state mesons and baryons. In contrast, for $N(1535)$ one needs 
some components other than hadronic composites, such as quark originated states. 
This twofold character of $N(1535)$, meson cloud and valense quark,  
can be seen also in the transition form factors of $\gamma^{*} N \to N(1535)$ 
(helicity amplitudes)~\cite{Ramalho:2012im}. 
In Ref.~\cite{Ramalho:2012im}, the $N(1535)$ transition form factors obtained 
by two different approaches, the chiral unitary approach~\cite{Jido:2007sm} and 
the spectator quark model~\cite{Ramalho:2011ae,Ramalho:2011fa}, are compared. 
The quark model calculation of the transition form factors tells that the $F^{*}_{1}$
form factor is produced well, while the $F^{*}_{2}$ amplitude is overestimated
in higher $Q^{2}$, where the quark model may be applicable. It is very interesting
that the meson cloud components calculated by the chiral unitary model compensates 
the disagreement of the quark model calculation as seen Fig.~\ref{fig:FF}.

In the chiral unitary approach, one can calculate the diagonal component of the electromagnetic form factor, and in Ref.~\cite{Sekihara:2010uz} the first moment
of the $\Lambda(1405)$ form factor, which corresponds to the spatial radius if 
the particle is stable, and it has been found that  $\Lambda(1405)$ has a 
substantially large size compared to the typical hadron.  The ``radius''
of unstable particles are obtained as a complex number and its interpretation 
should be done carefully.

\section{Kaonic few-body systems}

%\begin{figure}[htb!] 
%\centering 
\begin{wrapfigure}[10]{r}{0.45\textwidth}
\includegraphics[width=0.45\textwidth,bb=0 0 794 340]{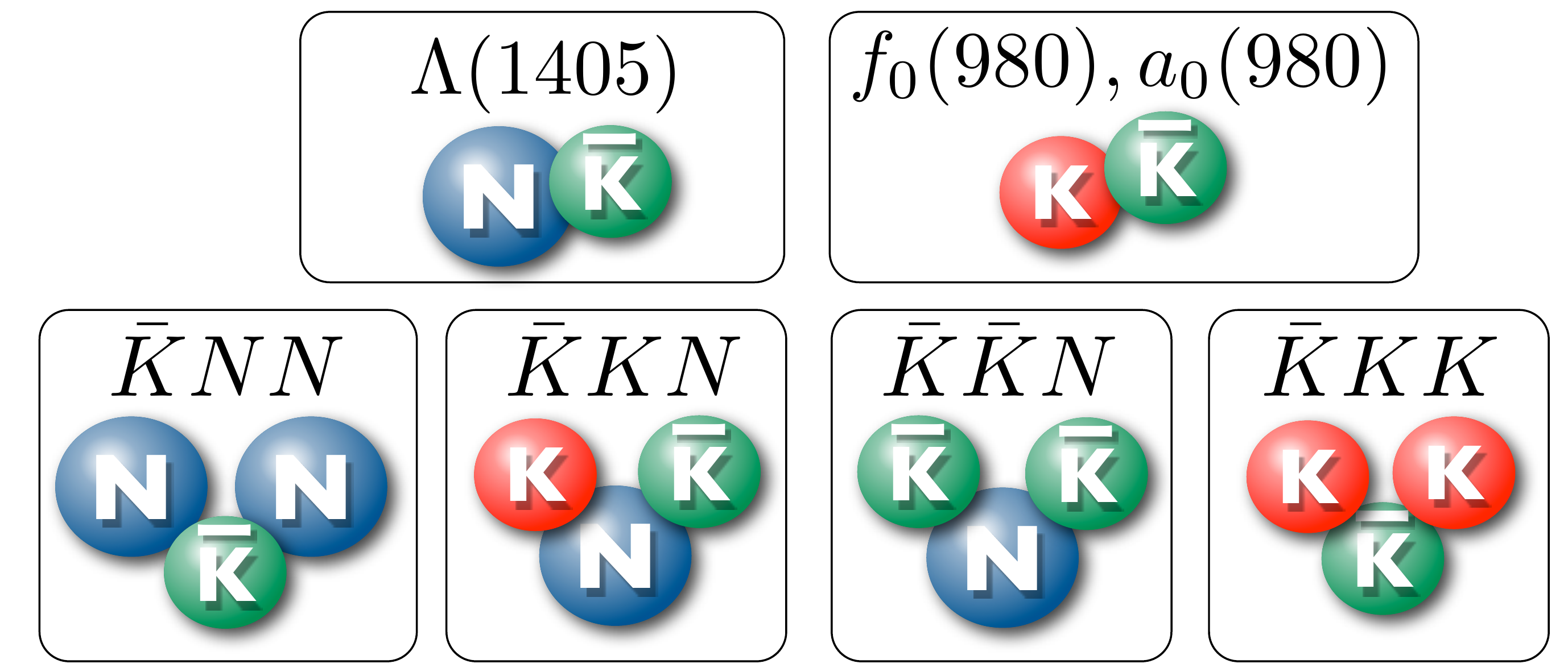} 
\caption{Family of kaonic few-body states. \label{fig:family} }
\end{wrapfigure}
%\end{figure} 
%--------------------------------------------------------------------- 

Let us consider hadronic composite states with kaon and nucleon constituents
as an extension of  the $\Lambda(1405)$ resonance of a quasibound state 
of $\bar KN$.
For hadronic composite states, pion has a too light mass to form bound states 
with other hadrons by hadronic interaction, since the pion kinetic energy in
a hadronic composite system  
overcomes attractive potential energy.
In contrast, kaon has a unique feature in hadronic composite state. 
The mass of kaon is so moderately heavy 
that kaon kinetic energy in hadronic bound systems can be smaller.
In kaonic few-body systems, hadronic molecular states are unavoidably resonances 
decaying into pionic channels.
The interaction of kaon as a Nambu-Goldstone boson is described well chiral effective theory.
It suggests that 
$s$-wave interactions in the $\bar KN$ and $\bar KK$ channels are attractive 
enough to form two-body quasibound states. 
The Tomozawa-Weinberg interaction is a driving force
of the hadronic molecular system. It is very interesting that the strength of the 
Tomozawa-Weinberg interaction is fixed by SU(3) flavor symmetry and 
$K$ and $N$ are classified into the same state vector in the SU(3) 
octet representation. Therefore, 
considering also that $K$ and $N$ are a similar mass, one finds that 
the fundamental
interactions in $s$-wave are very similar in the $\bar KK$ and $\bar KN$ channel.
Consequently these channels with $I=0$ have quasibound states of 
$\bar KK$ and $\bar KN$ with a dozen MeV binding energy. 
This similarity between $K$ and $N$ is responsible for systematics of three-body
kaonic systems, $\bar KNN$, $\bar KKN$, $\bar K \bar K N$ and $\bar KKK$, as shown 
Fig.~\ref{fig:family}. Further few-body systems with kaons and nucleons have been studied, for instance, in $\bar KNNN$ and $\bar K \bar K NN$ systems~\cite{Akaishi:2002bg,Barnea:2012qa}.

The $\bar KKN$ and $\bar K \bar KN$ states with $I=1/2$ and $J^{p}=1/2^{+}$, 
which are $N^{*}$ and $\Xi^{*}$ resonances, respectively, were studied 
first in Refs.~\cite{Jido:2008kp,KanadaEnyo:2008wm} 
with a single-channel non-relativistic potential model. The $\bar KKN$
system was found to be bound with 20 MeV binding 
energy~\cite{Jido:2008kp}, and later  was investigated in a more 
sophisticate calculation~\cite{MartinezTorres:2008kh,MartinezTorres:2010zv} 
based on a coupled-channels Faddeev method developed in 
Refs.~\cite{MartinezTorres:2007sr},
in which a very similar state to one obtained in the potential model was found. 
It was found also in a fixed center approximation
of three-body Faddeev calculation~\cite{Xie:2010ig}. The $\bar KKN$ state is
essentially described by a coexistence of $K \Lambda(1405)$ and 
$f_{0}(980)N$~\cite{Jido:2008kp}. An experimental search for  $\bar KKN$ 
was discussed in Ref.~\cite{MartinezTorres:2009cw}.
The $\bar KKK$ state with $I=1/2$ and $J^{p}= 0^{-}$, being an excited state of kaon,
was studied in a two-body $f_{0}K$ and $a_{0}K$ dynamics~\cite{Albaladejo:2010tj},
in the three-body Faddeev calculation~\cite{Torres:2011jt} and in the non-relativistic 
potential model~\cite{Torres:2011jt}. The three-body Faddeev calculation was done 
in coupled-channels of $\bar K KK$, $K \pi\pi$ and $K\pi\eta$ and a resonance 
state was found at 1420 MeV, while the potential model suggested a quasi bound
state with a binding energy 20 MeV. This state is essentially described by 
the $\bar KKK$ single channel and its configuration is found to be mostly $f_{0}K$. 
Experimentally, Particle Data Group tells that there is a excited kaon around 1460 
MeV observed in $K\pi\pi$ partial-wave analysis, although it is omitted 
from the summary table. 

In the potential model calculations of the $\bar KKN$ and $\bar KKK$ states,
it was found that the root mean-squared radii of these systems are as
large as 1.7 fm, which are similar with the radius of $^{4}$He.
The inter-hadron distances are comparable with an average 
nucleon-nucleon distance in nuclei. It was also found that the two-body
subsystems inside the three-body bound state keep their properties in 
isolated two-body systems. These features are caused by weakly binding of the 
three hadrons.

\section{Conclusion}

There should be hadronic composite states in which hadrons including 
mesons are constituents of the state in hadron spectrum as one of the 
existence forms of hadrons. These states have spatially larger sizes than the 
typical confinement range, because the driving force of the hadronic 
composite state is inter-hadron interaction which is out of the confinement
range. 

The $\Lambda(1405)$ resonance is one of the strong candidates 
of the hadronic composite states. The peculiarity  of the $\Lambda(1405)$ 
resonance is not only being a hadronic composite object but also 
being composed of two resonance states. These two states stems 
from the presence
of two attractive channels in fundamental meson-baryon interactions.
Eventually the observed $\Lambda(1405)$ resonance is composed by 
two pole states.
One of the states is a quasi-bound state of $\bar KN$ located at around
1420 MeV and dominantly couples to the $\bar KN$ channel. Thus, this
is the relevant resonance for the kaon-nucleus interaction.
The double pole structure of $\Lambda(1405)$ can be confirmed experimentally 
by observing $K^{-}d \to \Lambda(1405) n$, in which 
$\Lambda(1405)$ is produced selectively by the $\bar KN$ channel
and the  peak position appears around 1420~MeV.

The idea that $\Lambda(1405)$ is a quasibound state of $\bar KN$
can be extended systematically to further few-body states with kaons 
like $\bar KNN$, $\bar KKN$ and $\bar KKK$ having dozens MeV 
binding energy. 
In these states, a unique role of kaon is responsible for the systematics 
of the few-body kaonic states. Kaon has a half mass of nucleon and 
a very similar coupling nature to nucleon in the $s$-wave chiral interaction. 
This leads to weakly bound systems within the hadronic interaction range.
The hadronic composite state is a concept of weakly binding systems 
of hadron constituents. If a resonance state has a large binding energy
measured from the break-up threshold, coupled-channel effects, like
$\pi\Sigma$ against $\bar KN$, and/or shorter range quark dynamics
should be important for the resonance state. In such a case the hadronic
compostite picture is broken down, and one should take into account 
coupled channels contributions and quark dynamics. 

The hadronic composite configuration is a complemental picture 
of hadron structure to constituent quarks, which successfully 
describe the structure of the low-lying baryons in a simple way.
Strong diquark configurations inside hadrons can
be effective constituents~\cite{Kim:2011ut}, 
and mixture of hadronic molecular and quark originated states
is also probable in some hadronic resonances~\cite{Nagahiro:2011jn}.
The hadronic molecular state has a larger
spacial size than the typical low-lying hadrons. 
In heavy ion collision, coalescence of hadrons to produce 
loosely bound hadronic molecular systems is more probable 
than quark coalescence for compact multi-quark 
systems~\cite{Cho:2010db,Cho:2011ew}. 
Thus, one could extract the structure of hadrons by observing  
the production rate in heavy ion collisions.

\begin{acknowledgements}
The author would like to acknowledge his collaborators of this work, T.~Hyodo, A.~Hosaka, A.~Mart\'{i}nez~Toress, Y.~Kanada-En'yo,  E.~Oset, T.~Sekihara, J.~Yamagata-Sekihara, J.A.~Oller, A.~Ramos, U.-G.~Meissner, M.~Doring, G.~Ramalho, K.~Tsushima.
This work was partially supported by the Grants-in-Aid for Scientific Research (No. 24105706), and done in part under the Yukawa International Program for Quark-hadron Sciences (YIPQS).
\end{acknowledgements}

%% BibTeX users please use one of
%\bibliography{FB20_Jido}   % name your BibTeX data base

%% Non-BibTeX users please use
%\begin{thebibliography}{3}
%%
%% and use \bibitem to create references. Consult the Instructions
%% for authors for reference list style.
%\bibitem{Ref1}
%Hamburger, C.: Quasimonotonicity, regularity and duality for nonlinear systems 
%of partial differential equations. Ann. Mat. Pura. Appl. 169, 321-354 (1995)
%
%\bibitem{Ref2}
%Sajti, C.L., Georgio, S., Khodorkovsky, V., Marine, W.: 
%New nanohybrid materials for biophotonics. Appl. Phys. A (2007). 
%doi:10.1007/s00339-007-4137-z
%\bibitem{Ref3}
%Geddes, K.O., Czapor, S.R., Labahn, G.: Algorithms for Computer Algebra. Kluwer, Boston (1992)
%
%% Format for book chapter
%\bibitem{Ref4}
%Broy, M.: Software engineering  from auxiliary to key technologies. 
%In: Broy, M., Denert, E. (eds.) Software Pioneers, pp. 10-13. Springer, Heidelberg (2002)

% etc

%\end{thebibliography}

\end{document}